\documentstyle[12pt]{article}
\begin{document}
\newcommand{\vi}{\vec{b}}
\newcommand{\vj}{\vec{b}'}
\renewcommand{\theequation}{\thesection.\arabic{equation}}
\setcounter{page}{2}
\begin{titlepage}
\renewcommand{\thefootnote}{\fnsymbol{footnote}}
\begin{flushright}
\large{CTP-2390} \\
\large{FAU-TP3-94/5} \\
\large{December 1994}
\\
\end{flushright}
\vspace*{0.4cm}
\begin{center} \LARGE
{\bf  Signatures of Confinement \\ in Axial Gauge QCD}
\end{center}
\vspace*{0.4cm}
\begin{center}

{\bf F. Lenz}

\vspace*{0.1cm}

{\em Institute for Theoretical Physics, University of
Erlangen-N\"{u}rnberg,
\\
Staudtstr. 7, 91058 Erlangen, Germany$^{\,\,\dagger}$
\\
and
\\
Center for Theoretical Physics, Laboratory for Nuclear Science,
\\
MIT,
Cambridge, MA 02139, USA}

\vspace*{0.2cm}

{\bf  E. J. Moniz}

\vspace*{0.1cm}

{\em Center for Theoretical Physics, Laboratory for Nuclear Science,
\\
MIT,
Cambridge, MA 02139, USA$^{\,\dagger}$
\\
and
\\
Institute for Theoretical Physics, University of
Erlangen-N\"{u}rnberg
\\
Staudtstr. 7, 91058 Erlangen, Germany}

\vspace*{0.2cm}

{\bf M. Thies}

\vspace*{0.1cm}

{\em Institute for Theoretical Physics, University of
Erlangen-N\"{u}rnberg
\\
Staudtstr. 7, 91058 Erlangen, Germany}

\end{center}

\vspace*{1cm}

\centerline{\Large\bf Abstract}
\vspace*{0.2cm}

A comparative dynamical study of axial gauge QED and QCD is presented.
Elementary excitations associated with particular field
configurations are investigated. Gluonic excitations
analogous to linearly polarized photons are shown to acquire
infinite energy. Suppression
of this class of excitations in QCD results from quantization of the
chromelectric flux and is interpreted as a dual Meissner effect, i.e. as
expulsion from
the QCD vacuum of chromo-electric fields which are constant
over significant distances.
This interpretation
is supported by a comparative evaluation of the interaction energy of static
charges in the axial gauge representation of QED and QCD.
\vskip 2.0cm

\noindent
$^{\dagger}$ permanent address
\end{titlepage}
\addtocounter{footnote}{0}

\newpage

\section{Introduction}

 Confinement of colored objects is one of the fundamental properties
of  the strong interaction.
 Apart from the obvious importance for the spectrum of observed hadrons
in the absence of free gluons and quarks, confinement is quite
likely responsible for other more general properties such as the existence
of  Regge
trajectories and the spontaneous breakdown of chiral symmetry. It is believed
that color confinement is only one of the possible phases associated
with the strong
 interaction
which at  finite temperature and possibly at finite baryon density
undergoes a transition to an unconfined phase.
Theoretical evidence based on lattice
gauge calculations strongly suggests that QCD indeed exhibits the
phenomenon of
confinement and the transition to an unconfined phase at finite
temperature.
Significant experimental
efforts are underway  to establish the existence and to investigate
the properties of  the unconfined phase.

 Despite many efforts, a generally  accepted analytical
explanation or  qualitative description of confinement within the
framework of QCD is still missing. The only  theoretical model
of  a confining theory, which has a well defined derivation from  QCD,
is provided by
the strong coupling
limit of  lattice QCD. Unfortunately, this strong coupling limit
does not distinguish
QED and QCD as far as confinement is concerned. These two theories
supposedly develop
their characteristic differences only in a phase transition, which as a
function of the coupling constant is known to occur for  QED \cite{GUTH}
and for which  no evidence is found in lattice QCD calculations.

Canonical ``gauge-fixed'' formulations
of QCD \cite {BJOR,JACK} constitute a complementary approach to
the issue of confinement.
We will show how, within this framework, non-abelian properties of QCD
provide strong signatures of confinement. Distinguishing elements of
QED and QCD remain manifest.
A variety of  gauge fixed formulations of  QCD have already been
studied  (cf. \cite{BJOR}, \cite{CHLE} -- \cite{BALU})
with the aim of clarifying the structure of QCD.
The general strategy
of all these efforts consists in explicitly resolving the constraint
equations of QCD,
such as the Gauss law in the Weyl-gauge, and  thereby eliminating
the redundant
variables from the QCD Hamiltonian.  After this elimination, the
Hamiltonian is, by
construction, formulated in terms of gauge invariant variables
(leaving aside the issue of global, residual symmetries).
The resulting gauge fixed formulation of the dynamics is
accessible to approximate treatments which are fully
compatible with the underlying
local gauge symmetry. Without gauge fixing, approximations generally
violate gauge invariance and are therefore not
appropriate for investigations of issues like confinement, where
exact local color
conservation must be particularly important.

The paradigm of a gauge fixed canonical formalism is provided by the
Coulomb-gauge
representation of QED. Here, by explicitly resolving  Gauss's law  the
longitudinal components of the vector potential are eliminated. The resulting
description in terms of transverse vector fields is explicitly
gauge invariant irrespective of the approximations involved in application
of the formalism to specific dynamical problems. This gauge choice has played
a very important role also in the development of gauge fixed
formulations of QCD
\cite{BJOR,CHLE,GESA,CREU}.
However while in QED  the Coulomb-gauge  is singled out as the gauge
in which static charges do not radiate, gluons and color spin
remain coupled in QCD in any gauge.
 Moreover, the choice of radiation gauge variables
does not conform with the angular momentum algebra of the QCD Gauss law
operators, thereby causing significant technical disadvantages.
As a consequence,
the Hamiltonian of QCD in the Coulomb-gauge representation cannot be
constructed
explicitly; its calculation involves inversion of the  Fadeev-Popov
operator which can be performed only  perturbatively. Therefore, within
the canonical formalism, the Coulomb-gauge
has been particularly useful in perturbative studies
of, for example, the running coupling constant
\cite{BJOR,KHRIP,GOTT,DREL} or the
small volume limit of QCD \cite{LUES}. However dynamical
calculations within the
Coulomb-gauge are of great technical complexity \cite{BAAL}.

In the absence of compelling  physics reasons, the choice of gauge
has been  dictated  in most studies  by
formal and technical considerations. We are only at the beginning of
a development where common properties of the  resulting  quite different
formulations of QCD, such as the appearence of  centrifugal barriers in the
electric field energy \cite{LNT,BAFH,HAJO}, are being
recognized and their dynamical role is being understood. At this point it
appears
unlikely that
the non-perturbative phenomena of QCD are most easily described in
one unique  gauge. Given the complexity of  the Hamiltonian in all
the gauge fixed formulations,   it seems appropriate to
attempt to identify in a first step those
degrees of freedom whose dynamics gets particularly simplified  in a special
gauge and concentrate the studies on their  dynamics. We present here
the results of
such an investigation in the framework of the axial gauge representation of
QCD.

In the axial gauge representation, the ``simple'' degrees of freedom are
linearly polarized plane waves. This simplification arises by  identifying
one of the
coordinate axes (the 3-axis in the following)  with the spatial direction
of the
electric field and furthermore, in exploiting local gauge invariance, by
identifying the color
3-axis of  SU(2) QCD with the color orientation of the
electric field. It can be shown \cite{LNT} that, in implementing the
Gauss law, all other field configurations with electric fields pointing
in the spatial 3-direction have a longitudinal component and can therefore
be eliminated.  The simplification of the dynamics occurs for both  QED and
QCD. In order to control infrared difficulties the system
is enclosed in a box with periodic boundary
conditions.

The main part of our studies will be concerned with the elementary excitations
of these ``simple'' degrees of freedom. In QED, these elementary excitations
are nothing else than photons described in this particular gauge by a single
field. In QCD, the corresponding elementary excitations will be seen to not
exhibit the standard dispersion equation for massless particles; rather
these excitations are found to be frozen  in the infinite volume limit.
We shall interpret the disappearence of this class of  excitations from the
spectrum
as an indication of the dual Meissner effect: the QCD vacuum apparently
does not support color-electric fields which remain constant in a certain
direction over ``large" distances. The parallel treatment of QED and QCD
in analogous gauges is very important for identifying those
elements of QCD which make it differ so significantly from QED. Unlike
lattice theories, the canonical description
is not, by assumption, formulated in terms of compact variables; a restriction
in the range of certain dynamical variables occurs rather as a consequence of
modifications
in the electric field energy which are typical for non-abelian theories.
As another application of the simplified dynamics of plane wave excitations
we consider the static quark-antiquark interaction. Again, a proper choice
of coordinates and variables allows us to simplify significantly the dynamics
and, under additional assumptions, derive within the canonical gauge fixed
formalism the strong coupling result of lattice gauge theories.

\section{Formalism of axial gauge QCD}

 Our investigation starts with the QCD Hamiltonian in the axial
 gauge. For SU(2) color and in the presence of static charges,
the  Hamiltonian density is given by \cite{LNT}
\begin{equation}
{\cal H}  =   \mbox{tr} \left[  \vec{E}_{\bot}^{2}  +
 \vec{B}^{\,2} \right]+\frac{1}{2}{\cal E}_{3}^{2}+
\frac{1}{2 L^{2}} e_{3}^{\dagger}\left(\vec{x}_{\perp}\right)
e_{3}\left(\vec{x}_{\perp}\right)+ \frac{1}{2}
\vec{\eta} \, ^{2}\left(\vec{x}_{\perp}\right) \ .
\nonumber
\label{1}
\end{equation}
In the axial gauge representation of QCD, the degrees of
freedom are the (three color projections of the) perpendicular
components of the gauge fields $\vec{A}_{\perp}$ and their
conjugate electric fields $\vec{E}_{\perp}$. (We use an unconventional
sign for the electric field, so that $\vec{A}$ and $\vec{E}$ satisfy
standard commutation relations of coordinates and momenta.)
One of the cartesian
components of the fields, the 3-component, has been eliminated.
Naive gauge fixing gives rise to infrared problems; these  can be
avoided by enclosing  the system  in a box (length $L$) and imposing
periodic boundary conditions. In this way, two-dimensional, i.e.
$x_{3}$-independent, color neutral gauge fields
$a_{3}\left(\vec{x}_{\perp}\right)$ and their conjugate  electric
fields $e_{3}\left(\vec{x}_{\perp}\right)$ remain as unconstrained
degrees of freedom. In turn, the two-dimensional neutral,
longitudinal components of $\vec{A}_{\perp}$ and $\vec{E}_{\perp}$ are
identified as dependent degrees of freedom and have been
eliminated, i.e.,
\begin{equation}
\mbox{div}_{\perp}\int_{0}^{L}dx_{3}\vec{E}_{\perp}^{3} \left(\vec{x}
\right)=0    \ .
\label{2}
\end{equation}
(Note that the canonical commutation relations are correspondingly
modified.)
The relation between   magnetic  and  gauge fields is the
standard one,
\begin{eqnarray}
B_{3} & = & \partial_{1} A_{2} - \partial_{2} A_{1} -ig \left[
A_{1}, A_{2} \right]  \ ,
\nonumber \\
B_{i} & = & \epsilon_{ik3}\left(\partial_{k} a_{3}\tau^3/2 - \partial_{3}
A_{k} -ig \left[ A_{k}, a_{3}\tau^{3}/2 \right]\right)
\ \ \ \ \ \ (i=1,2) \ .
\label{4}
\end{eqnarray}
Except for  the modifications by the two-dimensional fields, the
contributions of the perpendicular degrees of freedom  to electric
and magnetic field energy are as usual; the second term in
eq. (\ref{1}) is just the contribution of the 3-component of the
electric field
\begin{equation}
\frac{1}{2}{\cal E}_{3}^{2}=\frac{1}{L^{2}}\int_{0}^{L} dz_{3}
\int_{0}^{L} dy_{3} \sum_{p,q,n}\left(1-
\delta_{p,q}\delta_{n,0}\right)
\frac{ G_{\bot pq}^{\dagger}(\vec{x}_{\bot}, z_{3}) G_{\bot
pq}(\vec{x}_{\bot},y_{3})}{\left[ \frac{2\pi n}{L} +
g\left(p-q\right)a_{3}(\vec{x}_{\bot}) \right]^{2}}
e^{i2\pi n(z_{3}-y_{3})/L}  \ ,
\nonumber
\label{5}
\end{equation}
which, by resolving Gauss's law, is given in terms of the other
degrees of freedom,
\begin{equation}
G_{\perp}\left(\vec{x}\right)=\vec{\nabla}_{\perp}\vec{E}_{\perp}
\left(\vec{x}\right)+g\epsilon^{abc}\frac{\tau^{a}}{2}\vec{A}^{b}
_{\perp}\left(\vec{x}\right)\left(\vec{E}^{c}_{\perp}\left(\vec{x}
\right)-\vec{\eta} \left(\vec{x}_{\perp}\right)\delta_{c,3}\right)
+ g\rho^{m}(\vec{x})  \ .
\label{6}
\end{equation}
We have omitted  the quark contributions in the Hamiltonian but
included for later application static color charges as described
by the density $\rho^{m}\left(\vec{x}\right)$.
The Hamiltonian density of eq. (\ref{1}) also contains the
two-dimensional, neutral color-electric  field $\vec{\eta}$
\begin{equation}
\vec{\eta}\,(\vec{x}_{\bot}) = \frac{g}{L} \vec{\nabla}_{\bot}
\int d^{3}y \, d\left( \vec{x}_{\bot} - \vec{y}_{\bot} \right)
\left( \epsilon^{3de} \vec{A}_{\bot}^{d}(\vec{x})
\vec{E}^{e}_{\bot}(\vec{x}) + \rho
^{3}_{m}\left(\vec{x}\right)\right)
\label{7}
\end{equation}
which is defined in terms of the two-dimensional Green function
\begin{equation}
d(\vec{z}_{\bot}) = - \frac{1}{L^{2}} \sum_{\vec{n} \neq \vec{0}}
\frac{1}{q_{n}^{2}} e^{i \vec{q}_{n} \vec{z}_{\bot}}  \ ,
\ \ \ \ \vec{q}_{n} = \frac{2 \pi}{L} (n_{1},n_{2}) \ .
\label{8}
\end{equation}
This ``electrostatic'' field  appears as a consequence of the
elimination of the neutral, longitudinal components from
$\vec{E}_{\perp}$ and $\vec{A}_{\perp}$ (cf. eq. (\ref{2})). Finally,
as a remnant
of the Gauss law constraint, periodicity of the fields requires the
neutral component of the charge to vanish in the space
of physical states $|\Phi\rangle$,
\begin{equation}
Q^{3}|\Phi \rangle = 2\int d^{3} x (G_{\perp})_{11}(x) |\Phi \rangle = 0 \ .
\label{8b}
\end{equation}

Further insight into the dynamics as described by the above QCD
Hamiltonian can be gained from a comparison with the Hamiltonian
of QED in the axial gauge
\begin{equation}
{\cal H}  = \frac{1}{2}\left[  \vec{E}_{\bot}^{2}  + \vec{B}
\,^{2}
\right]+\frac{1}{2}{\cal E}_{3}^{2}+ \frac{1}{2 L^{2}}
e_{3}^{2}\left(\vec{x}_{\perp}\right)+ \frac{1}{2}\vec{\eta} \,
^{2}\left(\vec{x}_{\perp}\right) \ .
\label{9}
\end{equation}
The same choice of the degrees of freedom as physical
($\vec{E}_{\perp}$) or constrained ($E_{3}$) ones  preserves the
similarity of the structure of the  underlying Weyl-gauge
Hamiltonians of QED and QCD. Obvious differences are
connected with the color structure, e.g. in the definition of the
QED magnetic fields ($g=0$ in eq. (\ref{4})), or in representing  the
3-component of the QED electric field in terms of the unconstrained
variables,
\begin{equation}
\frac{1}{2}{\cal E}_{3}^{2}=\frac{1}{2L^{2}}\int_{0}^{L} dz_{3}
\int_{0}^{L} dy_{3} \sum_{n}\left(1-\delta_{n,0}\right)
\frac{G_{\perp}(\vec{x}_{\bot}, z_{3})
G_{\perp}(\vec{x}_{\bot},y_{3})}{\left[ \frac{2\pi n}{L}
\right]^{2}} e^{i2\pi n(z_{3}-y_{3})/L} \ \nonumber
\label{10}
\end{equation}
with
\begin{equation}
G_{\perp}\left(\vec{x}\right)=\vec{\nabla}_{\perp}\vec{E}_{\perp}
\left(\vec{x}\right)+e\rho^{m}(\vec{x}) \ .
\label{11}
\end{equation}
In QED,  the electrostatic two-dimensional field  $\vec{\eta}$ is
generated by the density of the  charges which again are assumed
to be static,
\begin{equation}
\vec{\eta}\,(\vec{x}_{\bot}) = \frac{e}{L} \vec{\nabla}_{\bot}
\int d^{3}y \, d\left( \vec{x}_{\bot} - \vec{y}_{\bot} \right)
\rho^{m}\left(\vec{x}\right) \ .
\label{12}
\end{equation}
As in QCD, two-dimensional fields appear as a consequence of
implementing the Gauss law on a torus. Here, the physics reason is
particularly transparent. The electric fields
$e_{3}\left(\vec{x}_{\perp}\right)$ are purely transverse fields
and are therefore not part of the Gauss law constraint. These
fields together with the conjugate variables
$a_{3}\left(\vec{x}_{\perp}\right)$ describe photons polarized in
the 3-direction and propagating in the 1-2 plane. These degrees of
freedom cannot be eliminated while their counterparts, the purely
longitudinal components of $\vec{E}_{\perp}$, are completely
determined by the Gauss law and eliminated by the constraint
(\ref{2}).

\section{Elementary excitations }

The role of the two-dimensional degrees of freedom appearing in
both the QCD and QED axial gauge Hamiltonians can be discussed
from two quite different points of view.
On the one hand, their presence can be understood purely formally
to guarantee consistency of the formulation and provide proper
infrared behavior of the energy density. We note, for example, the
occurrence of $a_{3}\left(\vec{x}_{\perp}\right)$ in the
propagator defining
${\cal E}_{3}^{2}$ in eq. (\ref{5}).
Apart from this formal role  one might expect physically these
special two-dimensional degrees of freedom to contribute in a
thermodynamic sense negligibly to physical observables.
On the other hand, these two-dimensional fields generate
legitimate elementary excitations of the system.
In QED, any linearly polarized photon can, after an appropriate
choice of coordinates, be described by these lower dimensional
fields.
In this sense, the    degrees of freedom associated with
$a_{3}\left(\vec{x}_{\perp}\right)$ and
$e_{3}\left(\vec{x}_{\perp}\right)$ are of rather general nature,
and it is only  the peculiar choice of coordinates which
simplifies significantly their dynamics.

This simplification becomes explicit for QED by observing that
after an integration by parts in the magnetic field energy of eq.
(\ref{9}) (cf. also eq. (\ref{4})), the two-dimensional degrees of
freedom decouple from the remaining ones and are described by the
Hamiltonian
\begin{equation}
h=\int d^{2}
x\left[\frac{1}{2L}e_{3}^{2}\left(\vec{x}_{\perp}\right)+\frac{L}{
2}\left(\vec{\nabla}_{\bot}a_3(\vec{x}_{\bot})\right)^2\right] \ .
\label{13}
\end{equation}
We compare this Hamiltonian with the one for the corresponding
elementary excitations in QCD. Here, the color dynamics couple
the two-dimensional fields  to the perpendicular
degrees of freedom. In QED, the presence of charged matter would also induce
such a coupling.
In a first step we shall  neglect this coupling and keep \,-- with the
electric field energy --\, only the abelian contribution to the
magnetic field energy.
The resulting Hamiltonian for
$a_{3}\left(\vec{x}_{\perp}\right)$ of QCD is, by construction,
identical in structure with the QED one of eq. (\ref{13}) but for
the missing hermiticity of the QCD electric field operator
$e_{3}\left(\vec{x}_{\perp}\right)$ (cf. eq. (\ref{1})).
This difference has important consequences. The hermiticity
defect  arises since   SU(2) group elements ($W$),
actually loops around the torus along the 3-direction, have
been parametrized in terms of elements of the algebra ($a_3$),
\begin{equation}
W=\exp\left\{igL\vec{a}_{3}\vec{\tau}/2
\right\}=\cos\left(gLa_{3}/2\right)+i\vec{\hat{a}}_{3}\vec{\tau}
\sin\left(gLa_{3}/2\right)         \ .
\label{14}
\end{equation}
This is analogous to the
hermiticity defect of the radial momentum operator when using
polar coordinates and also leads here to modifications in the
kinetic energy. In the Schr\"odinger representation, expressing
the electric fields in terms of functional derivatives,
one finds
\begin{equation}
e_{3}^{\dagger}\left(\vec{x}_{\perp}\right)e_{3}\left(\vec{x}_
{\perp}\right)=-
\frac{1}{J\left(a_{3}\left(\vec{x}_{\perp}\right)\right)}
\frac{\delta}{\delta
a_{3}\left(\vec{x}_{\perp}\right)}J\left(a_{3}\left(\vec{x}_{\perp
}\right)\right)\frac{\delta} {\delta
a_{3}\left(\vec{x}_{\perp}\right)} \ .
\label{15}
\end{equation}
The Jacobian $J\left(a_{3}\right)$ is the Haar measure of SU(2),
\begin{equation}
J\left(a_{3}\left(\vec{x}_{\perp}\right)\right)=\sin
^{2}\left(\frac{1}{2} gLa_{3}\left(\vec{x}_{\perp}\right)\right)
\label{16}
\end{equation}
 and thus also appears in the volume element when evaluating matrix
elements in terms of wave functionals of
$a_{3}\left(\vec{x}_{\perp}\right)$.
As the transformation to polar coordinates  becomes singular when
attempting to define a direction for a vector of vanishing length,
 singularities  in the kinetic energy eq. (\ref{15}) arise for
$gLa_3=2\pi n$ in the parametrization of eq. (\ref{14}). It is convenient to
transform the kinetic energy into the standard form by defining, in
analogy to the Schr\"{o}dinger equation in polar coordinates, appropriate
``radial" wave functions. We introduce for this purpose a lattice in
the 1-2 plane (lattice constant $\ell$) and define the (rescaled) variables
at the lattice sites $\vi =\vec{x}_{\bot}/\ell$,
\begin{equation}
\varphi_{\vi}=\frac{1}{2}gLa_{3}(\vi\ell) \ .
\label{17}
\end{equation}
The radial wave functions $\tilde{\Phi}$ are defined in terms of the
original wave functions, which are the projections of the physical
states $|\Phi\rangle$ (cf. eq. (\ref{8b})) onto the field
eigenvectors $|\varphi \rangle$, in
the standard way
\begin{equation}
\tilde{\Phi}[\varphi]= \Phi[\varphi] \prod_{\vi}
|\sin (\varphi_{\vi})|   \ .
\label{18}
\end{equation}
In this way formally identical ``free" Hamiltonians of the two-dimensional
gauge degrees of freedom $\varphi$ for QED and QCD are
obtained
\begin{equation}
h=h^{e}+h^{m}\ ,
\label{19}
\end{equation}
with the electric and magnetic field energies given by
\begin{eqnarray}
h^{e} & = &  - \frac{g^{2}L}{8 \ell^{2}} \sum_{\vi}
\frac{\partial^{2}}{\partial \varphi_{\vi}^{2} } \ ,
\nonumber \\
h^{m} & = &  \frac{2}{g^{2}L}
\sum_{\vi,\vec{\delta}}\ \left(\varphi_{\vi +\vec{\delta}}-
\varphi_{\vi} \right)^{2}.
\label{20}
\end{eqnarray}
The fundamental vectors of the lattice $\left(1,0\right)$ and
$\left(0,1\right)$ are denoted by $\vec{\delta}$.
Although of the same structure, the Hamiltonian (\ref{19}) acts on
wave functions belonging to different spaces in QED and QCD.
The space of wave functions is restricted in QCD
by the constraint
\begin{equation}
\tilde{\Phi}[\varphi]=0 \quad \mbox{whenever} \quad \varphi_{\vi}=
n\pi \quad \mbox{for some}\  \vi \ .
\label{21}
\end{equation}
In the transformation of the Hamiltonian to ``radial" variables
no effective potential but an energy shift
\begin{equation}
E\rightarrow E-  \frac{g^{2} L^{3}}{8 \ell ^{4}}
\label{21a}
\end{equation}
appears, which has been suppressed in eq. (\ref{20}).

\subsection{Photons in the 1-2 plane}

In order to display the non-trivial dynamics described by the
Hamiltonian $h$ in conjunction with the constraint of
eq. (\ref{21}) we consider first the case of electrodynamics.
As is well known, in this case eigenstates and
energies are determined by discrete Fourier transformation
\begin{equation}
\varphi_{\vi}=\frac{1}{K}\sum_{\vec{k}}e^{i2\pi \vi \, \vec{k}/K}
\tilde{\varphi}_{\vec{k}} \ .
\label{24a}
\end{equation}
The number of degrees of freedom is $K^{2}$ with
\begin{equation}
K=\frac{L}{\ell}
\label{24}
\end{equation}
and limits the sum in eq. (\ref{24a}) to $|k_{1,2}| \leq K-1$.
In this way, the dispersion relation for the (lattice) photons,
\begin{equation}
\omega_{\vec{k}}^{2}=\frac{4}{\ell ^{2}}\sum_{\vec{\delta}}
\sin ^{2}\left(\frac{\pi\vec{\delta}\vec{k}}{K}\right)
\label{22}
\end{equation}
is obtained, with the continuum limit
\begin{equation}
\omega_{\vec{k}}^{2} \rightarrow \left(\frac{2\pi
\vec{k}}{L}\right)^{2}
\quad \mbox{for} \quad |\vec{k}|\ll K  \ .
\label{23}
\end{equation}
Thus the two-dimensional field
$a_{3}\left(\vec{x}_{\perp}\right)$ describes free photons
propagating in the 1-2 plane with  polarization in the 3-direction.
We also note that the dependence of the Hamiltonian on the
coupling constant is of no relevance, a simple rescaling
\begin{equation}
\varphi_{\vi}\rightarrow g L^{1/2} \varphi_{\vi}
\label{25}
\end{equation}
eliminates all but the dependence on the lattice spacing $\ell$.

In momentum space, the ground state wave function factorizes into
the contributions of the normal modes
\begin{equation}
\Phi[\tilde{\varphi}]=\prod_{\vec{k}}\left[\left(\frac{8\ell^{2}
\omega_{\vec{k}}}
{\pi g^{2}L}\right)^{1/2} \exp \left(-\frac{4 \ell ^{2}}{g^{2}L}
\omega_{\vec{k}}\tilde{\varphi}_{\vec{k}}\tilde{\varphi}_{-\vec{k}}
\right)\right]
\label{26}
\end{equation}
and thus describes  highly correlated zero-point motions in configuration
space.

\subsection{Jacobian and structure of the Hilbert space}

The nature of the elementary excitations as described by the
Hamiltonian (\ref{19}) is qualitatively altered by the presence of
the constraint (\ref{21}) on the wave function. This difference in
 dynamics is easily appreciated in a  mechanical interpretation.
While the  Hamiltonian (\ref{19}), (\ref{20}) describes
a two-dimensional system of
mass points which are coupled harmonically to the respective
nearest neighbors with the photons as the normal modes, the
constraint (\ref{21}) can be visualized as walls of an infinite
square well which limit strictly the amplitude of oscillations of
the individual mass points. Apparently, the character of the
normal modes of this system depends on the relative importance of
the harmonic nearest neighbor couplings and the constraining force.
The relevant parameter which controls the dynamics is the ratio
of constants multiplying electric and magnetic field energy,
\begin{equation}
\kappa=g^{2}\frac{L}{4\ell} \ .
\label{27}
\end{equation}
Unlike the case of electrodynamics, this ratio receives
significance by the constraint on the wave function which
prevents  the kinetic energy
from becoming  arbitrarily small for certain field configurations.

The presence of the constraint (\ref{21}) on the wave function
not only changes significantly the physical properties of the system
but also requires, in comparison
with the abelian case, a very different method of solution.
As a consequence of the constraint, the probability current
between regions of the wave functions separated by zeroes of the
Jacobian vanishes. Therefore the system defined by eqs. (\ref{19}) --
(\ref{21}) possesses an infinity of conserved charges
\begin{equation}
Q_{\vi}^{n}=\int_{n\pi}^{\left(n+1\right)\pi}
d\varphi\delta\left(\varphi-\hat{\varphi}_{\vi}\right)
\label{28}
\end{equation}
(here we have made explicit the operator character of
$\varphi_{\vi}$). These charges specify the probability of the system
to live in one of these intervals. In general, stationary states do not
exist for an arbitrary choice of these charges; they rather require each
of these charges to be non-vanishing only in one of the intervals. For
illustration we consider the case of two lattice sites.

In accordance with the constraint (\ref{21}) we decompose the wave function
\begin{equation}
\psi(\varphi_{1},\varphi_{2}) =
\sum_{n_{1},n_{2} = - \infty}^{+ \infty} \Theta_{n_{1}}(\varphi_{1})
\Theta_{n_{2}}(\varphi_{2}) \psi_{n_{1},n_{2}}(\varphi_{1},\varphi_{2}) \ ,
\label{29}
\end{equation}
where $\Theta_{n}$ denotes
the projection on the fundamental intervals
\begin{equation}
\Theta_{n}(\varphi) = \left\{ \begin{array}{l}
1 \ \ \mbox{for} \ \ n \pi \leq \varphi \leq (n+1)\pi \\
0 \ \ \mbox{otherwise} \end{array} \right.  \ .
\label{30}
\end{equation}
Most important, the wave function constraint decouples the solution of
the Schr\"{o}dinger equation in the different intervals; consequently
eigenvalues
and eigenfunctions can be obtained by choosing the wave function to be
non-vanishing in one of the intervals only, i.e.,
\begin{equation}
\psi_{n_{1},n_{2}}(\varphi_{1},\varphi_{2}) \neq 0 \quad \mbox{only if}
\quad n_{1}=n_{1}^{0}, \quad n_{2}=n_{2}^{0} \ .
\label{31}
\end{equation}
The complete spectrum is determined by variation of $ n_{1}^{0}$ and
$n_{2}^{0}$. In general, the potential energy and therefore the spectrum
depends on the choice of the intervals and therefore the system cannot coexist
in different intervals. When symmetries are present however, the wave function
corresponding to identical eigenvalues in different intervals can be
arbitrarily distributed over these intervals. For example, this happens
as a consequence of the displacement symmetry of the Hamiltonian
(\ref{19}), (\ref{20}); a global shift of the intervals
\begin{equation}
n_{\vi}^{0}\rightarrow n_{\vi}^{0}+m
\label{32}
\end{equation}
does not change the spectrum and the wave functions can be linearly
superimposed. Further possibilities for combining wave functions
arise, for instance, by the following more symmetric form of the discretized
magnetic field energy,
\begin{equation}
h^{m} \rightarrow   \frac{1}{g^{2}L} \sum_{\vi,\vec{\delta}}
 \left(1-\cos 2(\varphi_{\vi+\vec{\delta}}-
\varphi_{\vi}) \right) \ .
\label{33}
\end{equation}
The presence of symmetries makes the definition of the
eigenstates ambiguous; however the theory contains no operators which
would connect different intervals with each other and thereby be sensitive
to these ambiguities. In the following we shall restrict our
calculations to one definite (site-independent) choice of the fundamental
intervals,
\begin{equation}
Q_{\vi}^{n}=\delta_{n,0} \ .
\label{34}
\end{equation}
The site-independence guarantees that the state of lowest energy occurs
in this sector of the Hilbert space (with the parametrization (\ref{33})
of the discretized magnetic field energy, the spectrum is independent
of the choice of the fundamental intervals).

\subsection{Gluonic excitations in the 1-2 plane}

The {\em strong coupling limit}  ($\kappa\gg 1$) is physically most important.
We note that this limit is determined by the strength of the
dimensionless coupling constant $g$ relative to a negative power ($-1/6$) of
the number of degrees of freedom in the system (cf. (\ref{27})).
In the absence of a dimensionful
quantity, the continuum ($\ell \rightarrow 0$) and ``thermodynamic"
($L \rightarrow \infty$) limits are not distinguished. We therefore expect
the dependence of the coupling constant $g$ on the number of degrees of
freedom to be dictated by asymptotic freedom,
\begin{equation}
g^{2}(L/\ell) \propto 1/\ln \left(L/\ell\right)\ ,
\label{81}
\end{equation}
and  the thermodynamic limit to correspond to the
strong coupling limit.
In this limit, the electric field energy dominates  and the
stationary states are simply given by excitations of the degrees of freedom
at the individual lattice sites. The wave function of a stationary state can
be written as
\begin{equation}
\Psi_{\left[n\right]}[\varphi] = \prod_{\vi} \left[ \left(
\frac{2}{\pi}\right)^{1/2}
 \sin \left( n_{\vi}\varphi_{\vi} \right) \right]  \ ,
\label{35}
\end{equation}
with the energy eigenvalue
\begin{equation}
E_{n} = \frac{g^{2}L}{8\ell^{2}}\sum_{\vi} n_{\vi}^{2} \ .
\label{36}
\end{equation}
In particular, in the ground state all the individual degrees of freedom
are in the lowest $n=1$ state, and the value
\begin{equation}
E_{0} =  \frac{g^{2} L^{3}}{8 \ell ^{4}}
\label{37}
\end{equation}
is obtained for the ground state energy. This is just the negative of the
energy shift associated with the transformation to radial
coordinates (cf. eq. (\ref{21a})).
Thus the total ground state energy is actually zero and the
corresponding ground state wave function is an eigenfunction of the
modulus of the chromo-electric field operator
with vanishing eigenvalue (up to singular contributions arising
from possible discontinuities at $\varphi=n\pi$).
States of  lowest excitation   energy are obtained by exciting a
degree of freedom at one particular site into its first excited
state, with excitation energy
\begin{equation}
\Delta E = \frac{3}{8} \frac{g^{2} L}{\ell ^{2}} \ .
\label{38}
\end{equation}

Corrections to the strong coupling limit can be calculated perturbatively.
First order perturbation theory in $h^{m}$ (cf. eq. (\ref{20})) yields for the
ground state energy
\begin{equation}
E_{0} =  \frac{g^{2}L^{3}}{8\ell^{4}} + \frac{4L}{g^2\ell^2}\left(
\frac{\pi^{2}}{6}-1 \right) \ .
\label{39}
\end{equation}
The result justifies the perturbative treatment of the magnetic
field energy for sufficiently
large number of degrees of freedom $\kappa^{2}=g^{4} (L/4 \ell)^{2} \gg 1$.

Excited states of the system are highly degenerate in the strong coupling
limit. This degeneracy is lifted due to the magnetic coupling of the degrees
of freedom at different lattice sites. We consider the energetically lowest
excitation at the site $\vi$  described by the wave function
\begin{equation}
\chi_{\vi}[\varphi] = \left(\frac{2}{\pi}\right)^{1/2}
 \sin \left(2\varphi_{\vi} \right)\prod_{\vj (\neq \vi)}
\left[ \left(
\frac{2}{\pi}\right)^{1/2}
 \sin \varphi_{\vj} \right] \ .
\label{40}
\end{equation}
In the subspace of
these excitations,  the nearest
neighbor magnetic coupling becomes diagonal for the eigenstates of the
translation operator ($\vi \rightarrow \vi+\vec{\delta}$)
\begin{equation}
\tilde{\chi}_{\vec{k}}[\varphi]  =\frac{1}{K}\sum_{\vi}e^{-2i\pi\vi\,
\vec{k}/K}\chi_{\vi}[\varphi] \ .
\label{41}
\end{equation}
The calculation of the excitation energies is straightforward;
we find
\begin{equation}
\Delta E_{\vec{k}} = \frac{3}{8} \frac{g^{2} L}{\ell ^{2}}+\frac{64}{9\pi^{2}}
\frac{1}{g^{2}L}\left[\sum_{\vec{\delta}}\sin^2\left(\frac{\pi \vec{\delta}
\vec{k}}{K} \right) - 1 \right]
\label{42}
\end{equation}
corresponding to the state (\ref{41}) labeled by the momentum
$\vec{k}$.

The above  analysis of the strong coupling limit is the central
part of our investigations. We now present the physics implications
of the results. As already emphasized, the ground state of the system
is, in the strong coupling limit, an eigenstate of the electric
field operator with vanishing eigenvalue. This possibility for a ground
state with vanishing $x_{3}$-independent  electric fields arises
due to the Jacobian in the kinetic energy.
In QED such states are not normalizable and would entail infinitely
large fluctuations in the magnetic field energy. Thus the
structure of the vacuum concerning these $x_{3}$-independent
fields is very different in the abelian and non abelian theory.
The virial theorem applies to QED: magnetic and electric fields
contribute equally to each normal mode. Hence, the expectation value
$\langle \vec{E}^2-\vec{B}^2\rangle$ vanishes.
In QCD the Jacobian invalidates
equipartition. Chromoelectric $x_{3}$-independent
fields are absent; in turn, the fluctuations in the magnetic
field at different lattice sites are not correlated and the ground state
energy is due exclusively to these uncorrelated magnetic field fluctuations.
Similarly, the ``gluon condensate" is dominated by the magnetic
field contribution,
\begin{equation}
\langle \vec{E}^2-\vec{B}^2 \rangle = - \langle \vec{B}^2 \rangle < 0 \ .
\label{42a}
\end{equation}
It is interesting that even in the crude approximation of keeping only
one particular kind of gluonic degrees of freedom, this model
displays features which are reminiscent of the phenomenology of the
``magnetic QCD vacuum".

Concomitant with these qualitative differences in the structure
of the ground states is the very different nature of the elementary
excitations. Built on the highly correlated QED ground
state  the photons appear as  collective  excitations with excitation
energies vanishing in the  long-wavelength limit. In QCD,
the elementary excitations are localized in configuration space and
are due to formation of non-vanishing electric flux.
Intuitively, the expression (\ref{38}) for the excitation energy
can be seen as a result of a quantization of the chromoelectric flux.
In the absence of couplings to the other degrees of freedom, the
chromoelectric fields formed in the elementary excitations with lowest
excitation energy are located on just one transverse lattice site.
Thus the flux tubes are infinitely thin (i.e., the area is $\propto \ell^{2}$)
and extend over the whole system in the 3-direction. The chromoelectric
flux $\Phi_{E}$ associated with each component of the standing waves
(eqs. (\ref{40}), (\ref{41})) is quantized,
\begin{equation}
\Phi_{E}=ng \ .
\label{43}
\end{equation}
(Note that since the standing waves are not eigenstates of the electric
field operator, one has to decompose them into travelling wave components
in order to ``see" this flux quantization.)
Beyond the strong coupling limit, waves propagate through the medium
and transport the electric flux across a transverse plane. The spectrum
acquires a band structure with a band width $\propto 1/(g^2 L)$
 (cf. eq. (\ref{42})). Threshold energy
and  mass of the associated ``particles" are $\propto g^{2}L/\ell^2$
 and tend to infinity with increasing size of the system. Irrespective of
the value the transverse momentum, these elementary excitations never approach
the free gluon limit.

\section{Gluonic couplings}

Here we continue our investigation of the elementary excitations of QCD
as described by the two-dimensional
fields $\varphi_{\vi}$  (cf. eq. (\ref{17})).
Unlike in QED, where  these degrees of
freedom describe non-interacting photons in the absence of
charged matter, in QCD such a decoupling
does not occur. Here we study \,-- within perturbation theory --\,
the effect of the other degrees of freedom  of QCD
on the dynamics of  the  $ \varphi_{\vi}$. We shall show that the
effect of the infinity of  other degrees
of freedom is not sufficient to overcome the dominant  role  of the
kinetic energy of these particular two-dimensional
fields. The starting point of our investigation is the Hamiltonian
of eq. (\ref{1}) applied to
 SU(2) QCD.  With  the coupling constant $g$ treated as a small parameter,
the neutral gluons as described by
the conjugate pair of vector fields  $ \vec{A}_{\perp}^{3},\vec{E}_{\perp}
^{3}$ decouple from the other degrees of freedom and will not
be considered further.  At this point it is convenient to define
``charged'' gluon fields
\begin{equation}
 \vec{ \Phi }_{\perp}=\frac{1}{\sqrt{2}}\left(\vec{A}_{\perp}^{1}-
i\vec{A}_{\perp}^{2}\right) \quad , \quad
\vec{ \Pi }_{\perp} =\frac{1}{\sqrt{2}}\left(\vec{E}_{\perp}^{1}+
i\vec{E}_{\perp}^{2}\right) .
\label{44}
\end{equation}
The Hamiltonian  to be considered in the following,
\begin{equation}
h=h^{e}+h^{m}+ h'[\varphi] \ ,
\label{45}
\end{equation}
contains apart from the already discussed electric and abelian magnetic field
energy  the Hamiltonian $h ^{\prime}[\varphi]$ of the charged gluons  coupled
to the two-dimensional degrees of freedom $ \varphi_{\vi}$,
\begin{eqnarray}
h'[\varphi] & = & \int d^{3}x \left( \vec{\Pi}_{\bot}^{\dagger}(\vec{x})
\vec{\Pi}_{\bot}(\vec{x})+ \vec{\beta}^{\dagger}(\vec{x})
\vec{\beta}(\vec{x})\right)
\label{47}   \\
 & + & \frac{L}{4} \int d^{2} x_{\perp}\int_{0}^{L}dz \int_{0}^{L} dz'
\sum_{n}\vec{\nabla}_{\bot} \vec{\Pi}_{\bot}^{\dagger} (\vec{x}_{\perp},z)
\frac{e^{i2 \pi n
 (z-z')/L}}{(\pi n - \varphi(\vec{x}_{\perp}))^{2}}
\vec{\nabla}_{\bot} \vec{\Pi}_{\bot} (\vec{x}_{\perp},z') \ .\nonumber
\end{eqnarray}
The kinetic energy of the charged gluons receives contributions from the first
 and last term of the r.h.s. of the above equation. The last term also acts
as a centrifugal barrier on the two-dimensional degrees of freedom
$\varphi(\vec{x}_{\perp})$ (here, we have adopted a continuum notation
for these
variables, since most of the calculations can be carried through without
resorting to the lattice in the perpendicular direction).
The magnetic field $\vec{\beta}$ is defined as
\begin{equation}
\beta_{3}  =    \partial_{1} \Phi_{2} - \partial_{2} \Phi_{1} \quad , \quad
\beta_{2,1}  = \pm \left( \partial_{3}-  2i \varphi(\vec{x}_{\perp})/L \right)
\Phi_{1,2} \quad .
\label{48}
\end{equation}
As the centrifugal term, the  magnetic field energy contains a coupling between
 charged and neutral two-dimensional gluons via the covariant derivative
$(\partial_{3}-  2i \varphi /L)$.

The singular coupling (cf. eq. (\ref{47})) between the two-dimensional fields
$\varphi (\vec{x}_{\bot})$ and the charged gluons prevents straightforward
application of perturbation
theory. Expansion of the centrifugal term in eq. (\ref{47})
in terms of the variables $\varphi(\vec{x}_{\perp})$ yields non-integrable,
infrared
singularities.  Thus the following considerations are more generally required
for a systematic  expansion in $g$ when these two-dimensional fields with
their explicit non-perturbative dynamics are involved.

To define an appropriately decoupled zeroth order Hamiltonian
we introduce an $x_{\perp}$-independent external, neutral field $\chi$ which
represents the average equilibrium position of the $\varphi(\vec{x}_{\perp})$.
In this way a well defined, though parameter  dependent interaction
Hamiltonian
is introduced. Eventually, the external field $\chi$ will be treated as a
variational quantity, used to minimize the ground state energy. Formally, we
write
\begin{equation}
h' [\varphi] = h'[\chi] + \left( h'[\varphi] - h'[\chi] \right)
\label{50}
\end{equation}
and define the $\varphi$-potential energy resulting from
the coupling to the charged gluons as
\begin{equation}
h_{\rm int}= \langle \Phi_{\chi}^0|\left(h'[\varphi] -
h'[\chi]\right)| \Phi_{\chi}^0\rangle \ .
\label{51}
\end{equation}
The state vector $|\Phi_{\chi}^0\rangle$ denotes the $\chi$-dependent
ground state of the non-interacting charged gluons,
\begin{equation}
h'[\chi]|\Phi_{\chi}^0\rangle=E_{0}[\Phi]|\Phi_{\chi}^0\rangle \ .
\label{52}
\end{equation}

In the first step towards the determination of $h_{\rm int}$, we
diagonalize the free charged gluon Hamiltonian. Since $h'[\chi]$ is
quadratic in the charged gluon field operators, this diagonalization
is achieved by a Bogoliubov transformation and yields the result
\begin{equation}
h'[\chi] = \sum_{\lambda, \vec{n}} |\vec{p}(\vec{n})|
\left( \beta_{\lambda}^{\dagger}(\vec{n}) \beta_{\lambda}(\vec{n}) +
\alpha_{\lambda}^{\dagger}(- \vec{n}) \alpha_{\lambda}(- \vec{n}) +1 \right)
 \ . \label{53}
\end{equation}
The operators $\alpha_{\lambda}^{\dagger}(\vec{n}),\, \beta_{\lambda}^{\dagger}
(\vec{n})$ create the two (SU(2)) charged massless gluon  states with
polarization specified by $\lambda$ and momentum given by $\vec{k}(\vec{n})$,
\begin{equation}
\vec{k}(\vec{n})=\frac{2\pi}{L}\left(n_{1},n_{2},n_{3}\right) \ ,
\label{54}
\end{equation}
while their energies are  determined by the vector
\begin{equation}
\vec{p}(\vec{n})=\frac{2\pi}{L}\left(n_{1},n_{2},n_{3}-\chi/\pi\right)
\label{55}
\end{equation}
associated with the covariant derivative. The coupling to the external,
spatially constant neutral gluon field $\chi$ induces an asymmetry in the
energies of the two charge states. Furthermore, the independence of the
energies $|\vec{p}(\vec{n})|$ of the polarization $\lambda$ requires a
consistent treatment of both magnetic and electric coupling of $\chi$ to
the charged gluons. For vanishing $\chi$ and with the $n=0$ term in the
centrifugal term (cf. eq. (\ref{47})) disregarded, $h'[\chi]$ represents
(apart from
the corresponding two-dimensional fields) the free Hamiltonian for two types
of photons in the axial gauge representation. By the Bogoliubov transformation
this Hamiltonian gets transformed into the Coulomb-gauge representation and
thereby the non-locality in the electric field energy is eliminated
\cite{LNOT1}. Having determined the Bogoliubov transformed ground state
it is straightforward to calculate the separate contributions to the
ground state energy.
The color-electric contribution  $h ^{e}_{\rm int}$ to the potential
energy  $h_{\rm int}$  (cf. eq. (\ref{51}))  arises from the centrifugal
 barrier term in eq. (\ref{47}), the color-magnetic one $h ^{m}_{\rm int}$
 from the perpendicular components of the magnetic field,
\begin{equation}
h _{\rm int}=h ^{e}_{\rm int}+h ^{m}_{\rm int} \ .
\label{56}
\end{equation}
To evaluate these matrix-elements we replace at this point the corresponding
integrals over the perpendicular coordinates by sums and obtain
\begin{eqnarray}
h ^{e}_{\rm int} & = & \frac{\ell^{2}}{2 L^{2}} \sum_{\vi, \vec{n}}
\left\{ \left( \frac{k_{n3}- 2\varphi/L}{k_{n3}- 2 \varphi_{\vi}/L}
 \right)^{2}
-1 \right\} \frac{ k_{n \bot}^{2}}{\sqrt{p_{n3}^{2}+ k_{n\bot}^{2}}} \ ,
\nonumber \\
h ^{m }_{\rm int} & = & \frac{\ell^{2}}{2 L^{2}} \sum_{\vi, \vec{n}}
\left\{ \left( \frac{k_{n3}- 2\varphi_{\vi}/L}{k_{n3}- 2 \varphi/L}
  \right)^{2}
-1 \right\} \frac{2 p_{n3}^{2}+ k_{n \bot}^{2}}{\sqrt{p_{n3}^{2}+
k_{n\bot}^{2}}} \ .
\label{57}
\end{eqnarray}
(Here, we have used the shorthand notation $k_{n3}= k_3(\vec{n})$,
$k_{n\bot}=\vec{k}_{\bot}(\vec{n})$, etc.)
As could be expected, these energies are  divergent despite the
subtraction  of the most singular $\chi $ and
$\varphi _{ \vi }$-independent terms in the definition (\ref{51})
of $h _{\rm int} $.
Using a heat kernel regulator of the form
\begin{equation}
e^{- \lambda \sqrt{p_{n3}^{2} + k_{n \bot}^{2}}}
\label{58}
\end{equation}
and identifying the inverse momentum cut-off with the lattice spacing
\begin{equation}
\lambda = \gamma \ell  \ ,
\label{59}
\end{equation}
 we obtain  to leading order in $\ell/L$
\begin{eqnarray}
h ^{e}_{\rm int} & = & \frac{1}{2\pi \gamma ^{3}\ell} \sum_{\vi} \left\{
2 (\chi  - \varphi_{\vi} ) \cot \varphi_{\vi} + (\chi  -
\varphi_{\vi})^{2} \frac{1}{\sin^{2} \varphi_{\vi}} \right\} \ ,
\nonumber \\
h ^{m}_{\rm int}  & = & \frac{1}{2\pi \gamma ^{3}\ell }
\sum_{\vi} \left\{
2 (\varphi_{\vi}- \chi ) \cot \chi+ (\varphi_{\vi} - \chi )^{2}
\frac{1}{\sin^{2} \chi } \right\} \ .
\label{60}
\end{eqnarray}
Determination of the value of the external variable $\chi$
requires computation
of the expectation value of $h_{\rm int}$ in the unperturbed ground state
 ($n_{\vi}=1$ in eq. (\ref{35})). It is easily seen that
the interaction energy becomes minimal at
\begin{equation}
\chi=\pi/2 \ ,
\label{61}
\end{equation}
 i.e. at the midpoint of the intervals in which the $\varphi_{\vi}$
 degrees of
freedom move.
This choice also minimizes the total energy; the zero point energy in
$h'[\chi]$ (eq. (\ref{53})), though $\chi$-dependent, does not contribute to
leading order in $\ell/L$. Thus our final result for $h _{\rm int}$ reads
\begin{equation}
h_{\rm int}=\frac{1}{2\pi \gamma ^{3}\ell}\sum_{\vi}\left\{\left(
\pi/2-\varphi_{\vi}\right)^{2}\left(1+\left(\sin \varphi_{\vi}\right)^{-2}
\right)+2\left(
\pi/2-\varphi_{\vi}\right)\cot \varphi_{\vi}\right\} \ .
\label{62}
\end{equation}
The properties of the potential energy  $h_{\rm int}$ are directly
correlated
with the structure of the centrifugal term and the  non abelian magnetic field
 energy
in eq. (\ref{47}). This potential energy is local, i.e., it does
not connect degrees
of freedom at different lattice sites with each other. It therefore does not
significantly change the formation of waves across the transverse plane nor
their suppression in the  strong  coupling limit  ($\kappa\gg 1$). With the
centrifugal term, $h_{\rm int}$ is singular at
$\varphi_{ \vi }=0, \pi$.
Close to the origin the kinetic energy  of the ``radial''
motion  (cf. eq. (\ref{20})) is thus supplemented by a centrifugal barrier
\begin{equation}
-\frac{\partial^{2}}{\partial \varphi_{\vi}^{2}} \rightarrow
-\frac{\partial^{2}}
{\partial \varphi_{\vi}^{2}} +\frac{\pi\ell}{\gamma ^{3}g^{2}L}
\frac{1}{\varphi_{\vi}^{2}} \ .
\label{63}
\end{equation}
The presence of the centrifugal barrier forces  the full wavefunction
($\Phi $ in eq. (\ref{18})) to vanish
at the origin.  In addition to the centrifugal barrier, the potential
energy  (\ref{62})  exhibits
a repulsive Coulomb-like singularity.  The modification
of the ground state energy
due to $h_{\rm int}$  is given by
\begin{equation}
\Delta E_{0}=\frac{L ^{2}}{4\pi \left(\gamma  \ell\right)^{3}}
\left( 1+\pi^{2}/2\right) \ .
\label{64}
\end{equation}
Thus coupling to the charged gluons gives rise to a non-vanishing
and, in the
continuum limit,  divergent  energy density  (energy per unit area).
This is the
leading contribution in the continuum limit  (taking into account
the redefinition
of the  energy  in eq. (\ref{21a})). Nevertheless, the dominant operator
of the Hamiltonian (\ref{45})  remains the
electric field energy $h^{e}$ (cf. eq. (\ref{20})).

Finally we discuss
the result (\ref{61})
concerning the average field which, in view of  the singularities
at  $0,\pi$, is plausible.
This result  actually reflects a  symmetry of the exact Hamiltonian which
guarantees quite generally
that  the potential energy  of the $\varphi_{\vi}$ degrees
of freedom must
be stationary at $\pi/2$. As shown in \cite{LNT}, the
Hamiltonian  (\ref{1}) is
invariant under ``displacements'', ``central conjugations'',
and reversal of the color
3-axis. Following  arguments developed in the context of 1+1 dimensional QCD
\cite{LST}  we can combine these transformations to a relevant symmetry
transformation
which  changes the variables in the following way
\begin{eqnarray}
\varphi_{\vi} &\rightarrow  & -\varphi_{\vi} + \pi  \nonumber\\
 \vec{\Phi }_{\perp} \left(  \vec{x}\right)  & \rightarrow
& e^{-2i\pi x_{3}/L}
 \vec{\Phi }_{\perp}^{\,\dagger}  \left(  \vec{x}\right)
 \quad , \quad  \vec{\Pi }_{\perp} \left(\vec{x}\right)  \rightarrow
   e^{2i\pi x_{3}/L}
\vec{\Pi }_{\perp}^{\,\dagger}  \left(  \vec{x}\right) \nonumber \\
\vec{A}_{\perp} ^{\,3} \left(\vec{x}\right)  &\rightarrow  & -
\vec{A}_{\perp} ^{\,3}\left(\vec{x}\right) \quad ,\quad \qquad \ \  \
 \vec{E}_{\perp} ^{\,3} \left(\vec{x}\right) \rightarrow -
\vec{E}_{\perp} ^{3\,} \left(\vec{x}\right)  \ .
\label{65}
\end{eqnarray}
The invariance of the exact (eq. (\ref{1})) as well as of the
approximate (eq. (\ref{45}))
 Hamiltonian   under this transformation is easily verified.
After  integrating out the
charged gluons, the only remnant of the above symmetry is the
transformation of  the
variables $\varphi_{\vi}$. The symmetry therefore reduces to the
invariance of   $h_{\rm int}$
under a common reflection of  the variables $\varphi_{\vi}$
at $\pi/2$. This choice in turn
implies a  charge conjugation symmetric Hamiltonian  (cf. eqs. (\ref{53}),
(\ref{55})). As
the net effect  of  the coupling to the two-dimensional fields,
the spatially periodic
charged gluon fields acquire energies
$| \vec{p}\left(\vec{n}\right)|$   which actually correspond to
anti-periodic boundary conditions.

\section{Physics in axial gauge QCD }

On the basis of  the  results derived in the preceding sections, we shall
develop in this concluding section
 the physics picture of QCD as it presents itself
in the axial gauge. The essential result of our investigations
is the suppression of  plane wave type excitations,
i.e., excitations where the electric field is constant along a particular
spatial direction.
In our choice of coordinates and gauge,  we have identified
the spatial 3-axis with the direction in which the field does not vary
and have used the local gauge invariance to have the field pointed
 in the color 3-direction everywhere in space. With this choice of
coordinates, these particular degrees of freedom \,-- neutral, linearly
polarized gluons --\, are described by two-dimensional fields resulting in a
significantly simplified description of the dynamics. Such a simplification
is not  too surprising
considering the fact that in QED any linearly polarized photon,
with the corresponding choice of  coordinates and in the absence of
matter, is described by a non-interacting  two-dimensional scalar field.
In QCD the decoupling from the other degrees of freedom is not
complete, as happens in QED when coupled to charged matter.
A characteristic  difference to QED is that the gluon self interactions
give rise to a significant modification of the kinetic energy
(i.e.  electric field energy)
 of these particular degrees of freedom.  This modification is
a direct consequence of the projection onto physical states as obtained
 in the implementation of  the Gauss law. It is analogous to the modification
of the kinetic energy of  a quantum mechanical particle when  projected onto
states with definite angular momentum. As a result of this change,
 eigenstates to this modified kinetic energy operator of QCD
become normalizable, i.e., the electric flux  associated
 with these normalizable
 states gets quantized.
 Excitation energies  of  states with non-vanishing electric
 flux diverge linearly with the size of the system.
Up to this
point, the size of the system has been treated as a formal
parameter introduced
to properly define the theory in the infrared. Whenever  $L$
is large compared with the relevant intrinsic length scale \,-- $1/\Lambda$
after
renormalization or, phenomenologically, the size of hadrons --\, finite
size effects
should be negligible.
Success  of present lattice calculations actually indicates that
the size of the system may not have to be much larger than hadronic sizes.
In this sense, our results suggest that the QCD vacuum does not support
color-electric fields which are constant over distances larger than
hadronic sizes.
Excitation energies associated with such fields grow
with
the distance over which  such fields are constant.
This behavior is independent
of  their variation in transverse directions, i.e.,
independent of the associated
total momentum. Thus  finite excitation energies can result
only for structures
radically different from plane waves.  In summary,
color-electric fields  constant
over large distances are expelled from the QCD vacuum, i.e.  the QCD
vacuum exhibits the dual Meissner effect. This in turn implies that
magnetic fields
cannot be correlated  at points in space separated by
distances over which constant
electric fields cannot extend.

As an application of this picture, we consider the interaction between two
static color charges in comparison with the electrostatic interaction in the
Maxwell theory. In the axial gauge representation of QED,
the Coulomb-interaction
between static charges is not manifest, rather a linearly
rising potential appears in the
expression (\ref{10})  for the field energy. This contribution
is however only part of the
electrostatic interaction. In  the axial gauge representation
of QED (as in any
but the Coulomb-gauge) static charges couple to the radiation
field  as described by
the interference between  $ \vec{\nabla } _{\perp}  \vec{E}_{\perp}$ and
$ \rho^{m}$ in eq.
(\ref{10}).  It is easy to decouple the radiation field from
the charges by shifting
the electric field
\begin{equation}
 \vec{E}_{\perp} \rightarrow   \vec{E}_{\perp} + e \vec{\nabla } _{\perp}
 \frac{1}{\Delta}\rho^{m} \ .
\label{66}
\end{equation}
Thus by introducing  a  pair  of static charges (with net charge zero at a
distance $d$
from each other) into the QED vacuum,
the equilibrium position of the electric field changes. The radiation field
oscillates around
the electrostatic field generated by the static charges. For harmonic
oscillations, such a shift
in equilibrium position is irrelevant apart from a ``$c$-number'' change in
the zero point energy,
which is nothing else than  the electrostatic Coulomb-energy
\begin{equation}
U = \frac{e^{2}}{4\pi d} \ .
\label{67}
\end{equation}
 In QCD  a corresponding shift can also be performed; it decouples however
``static''
color charges and radiation field only perturbatively. Obviously, the dynamics
of the color spin of the static quarks necessarily remains coupled to the
gluons. On the other hand our above discussion suggests that
 the QCD vacuum resists such a shift in the
color-electric radiation field.
We have seen that the vacuum does not support electric fields which
have small variations
over large distances. Such fields are introduced however  in the
shift  (\ref{66}).
 In order to follow more closely how a linearly
dependent potential between static quark and antiquark arises in
axial gauge QCD we
assume \,-- beyond our above considerations --\, that the QCD vacuum is void
of any color-electric field. Introducing a static
quark-antiquark pair requires adjustment of the vacuum to account
for the correspondingly modified Gauss law. In a gauge fixed
formalism, this readjustment is enforced by the ``color-electrostatic''
field energy appearing in the Hamiltonian in the process of implementing
the Gauss law. In the axial gauge, the energy density of the electrostatic
field is given by the operator
$\frac{1}{2}\left({\cal E}_{3}^{2}+
\vec{\eta} \, ^{2}\right)$ in the Hamiltonian
of eq. (\ref{1}). With an appropriate choice of the coordinates, it is
possible to account for the necessary readjustment of the vacuum by
changes in one of the degrees of freedom only. For this purpose, the
coordinates have to be chosen such that quark and antiquark are located
on the
3-axis. In this case, we may assume that the color-electric fields
$\vec{E}_{\perp}(\vec{x})$ remain undisturbed and essentially zero.
 Also, due to the neutrality
condition which the static charges have to satisfy, the two-dimensional
neutral, color-electric fields $\vec{\eta}\left(\vec{x}_{\perp}\right)$
remain zero. Thus introduction of a static quark-antiquark system
affects  only the neutral two-dimensional
gluons $\varphi_{\vi}$ in the Hamiltonian (\ref{1}).
Furthermore, as we have seen,  degrees of freedom at
different lattice sites are decoupled in the
strong coupling limit. Only the neutral degree of freedom
$\varphi_{\vi_{0}}$ at the site $\vi_{0}$, corresponding to the
transverse coordinate of the
quark-antiquark position, is  coupled to the static charges. The Hamiltonian
describing this system of quantum mechanical degrees of freedom reads
\begin{equation}
\delta h =  - \frac{g^{2}L}{8 \ell^{2}}
\frac{\partial^{2}}{\partial \varphi_{\vi_{0}}^{2} } \ + u \ .
 \label{68}
\end{equation}
The coupling potential is the electrostatic field energy
${\cal E}_{3}^{2}/2$  (cf. eq. (\ref{5})),
\begin{equation}
u=\frac{g^{2}\ell^{2}L}{4}\int_{0}^{L} dz_{3}
\int_{0}^{L} dy_{3} \sum_{p,q,n}\left(1-
\delta_{p,q}\delta_{n,0}\right)\frac{
\rho^{m \,\dagger}_{pq}(\vi_{0}, z_{3})
\rho^{m}_{pq}(\vi_{0}, y_{3})}{\left[ \pi n +
\left(p-q\right)\varphi(\vi_{0}) \right]^{2}}
e^{i2\pi n(z_{3}-y_{3})/L}  \ ,
\label{69}
\end{equation}
which is given in terms of the static color charge density
\begin{equation}
\rho^{m}_{pq}\left(\vi,z\right)
= \frac{g}{2\ell^{2}}
\sum_{\alpha=q,\bar{q}}\left[
\psi^{\alpha \dagger}_{q}\left( \vi,z\right) \psi^{\alpha }_{p}
\left( \vi,z\right)
-\frac{\delta _{pq}}{2}\sum _{r=1,2} \psi^{\alpha \dagger}_{r}
\left( \vi,z\right)
\psi^{\alpha }_{r}\left( \vi,z\right)\right]\nonumber  \ .
\label{70}
\end{equation}
Here we have represented the color degrees of freedom of the static quarks
in second
quantized form. The operator $\psi^{\alpha \dagger}_{q}\left(
\vi,z\right)$
creates a
quark ($\alpha=q$) or antiquark ($\alpha=\bar{q}$) at the position
($(\vi\ell,z)$).
With our sequence of approximations we have reduced the calculation
of the static quark interaction energy to a problem of coupled quantum
mechanical degrees of freedom. The above Hamiltonian $\delta h$ is, after
redefinition of the coupling constant, identical with the Hamiltonian
describing the interaction
of static charges in SU(2) QCD in one space dimension. The spectrum
of this Hamiltonian
has been determined in \cite{ENSC} and yields for the interaction energy
of static color charges
\begin{equation}
U=\frac{3}{8}\frac{g^{2}d}{\ell^{2}} \ .
\label{71}
\end{equation}
This result agrees with the strong coupling limit of the SU(2) static
quark-antiquark
interaction in  Hamiltonian lattice gauge theory \cite{KOSU}. It is
quite satisfactory that approximations with similar physics content, the
common assumption of vanishing color-electric fields,  yield in the
two quite different formal approaches the same physical results. This is
particularly remarkable in view of the different ways in which the gauge
symmetry is treated.

For our derivation, the proper choice of coordinates has
been
instrumental. It is instructive to consider the problem with a choice of
the 3-axis
which does not coincide with the direction of the quark-antiquark
dipole moment. In this case, the color charges of quark and antiquark
contribute to the $y_{3},z_{3}$  integrations  for different values of the
perpendicular coordinates $\vi$. As a consequence, in the corresponding
$n=0$ terms, quark and antiquark contributions do not cancel. They rather
give rise to a centrifugal barrier with strength $\propto g^{2}L/\ell^{2}$.
In the continuum limit, an infinite interaction energy results.
Thus, with this choice of coordinates, the assumption
of vanishing color-electric fields $ \vec{E}_{\perp}$ is unattainable.
The system has to respond to the non-vanishing net charges arising in
the $x^{3}$ integrations and compensate these charges by color-electric
fields. Thereby a flux tube between quark
and antiquark is formed which, with such a choice of coordinates,
involves the complicated dynamics of  the coupled
$ \vec{E}_{\perp}$ fields   rather than the simple dynamics of the single
degree of freedom $\varphi_{ \vi_{0}}$.

Comparison of the result (\ref{71}) with the expression (\ref{38}) for the
energy associated with the lowest excitation of the two-dimensional
degrees of freedom $\varphi_{\vi}$ supports the arguments concerning
the quantization of electric flux
(cf. eq. (\ref{43})). We observe that unlike the lattice formulation, the axial
gauge representation does not introduce compact variables which by
construction lead to such a quantization of the electric flux. In axial gauge
QED no such phenomenon occurs. It is the appearence of the Jacobian
which forces the relevant QCD degrees of freedom not to be periodic but
rather to be constrained to a compact interval which is defined by
consecutive zeroes of the Jacobian. Thus most crucial properties
of the two-dimensional
degrees of freedom pertinent to the issues of confinement and the dual
Meissner effect are traced back to a fundamental difference
between QED and QCD which becomes manifest in the gauge fixed formulation.
In the lattice formulation of QCD, a similar dynamical role is played
by the Haar measure in the definition of the partition function. Our findings
are reminiscent of the disappearence of confinement when the Haar measure
is replaced by a constant \cite{PATR}.

In our discussion of the interaction energy of static quarks, a series of
approximations has been necessary  whose validity might be
difficult to assess quantitatively. On the other hand, our treatment
of the dynamics of the two-dimensional degrees of freedom
$\varphi \left(\vec{x}_{\perp}\right)$ has been quite straightforward and
therefore provides a novel access to the strong coupling
 limit of QCD. As the fundamental approximation in this approach,
the coupling of the neutral $\varphi \left(\vec{x}_{\perp} \right)$
to the charged gluon
fields has been neglected. With this coupling taken into account, the
flux tubes are expected to acquire finite transverse extension
and thereby the strong
coupling result for the excitation energy (\ref{38})  to be
determined by the  string tension
\begin{equation}
\frac{3g^{2}}{\ell^{2}} \rightarrow  \sigma  \ .
\label{72}
\end{equation}
We have performed a first step beyond the ``strong  coupling limit.''
The singular nature of the coupling  to the charged gluons prevents
straightforward application of perturbation theory and we have determined
in a variational calculation  the average effect of the charged gluons on the
$\varphi \left(\vec{x}_{\perp} \right)$  ground state. This improved
ground state
can serve as starting point of a systematic expansion. Calculation of  this
 lowest order correction reveals the presence of  two
different formal, small  parameters  involved in such an expansion. On the
one hand,  ordinary
perturbative treatment of the non-linearities of QCD   in both the
non-abelian part
of the magnetic field energy  (cf. eq. (\ref{4}))  as well as in the
non-abelian contribution to the
perpendicular Gauss law operator $G _{\perp} $  (\ref{6})  naturally requires
\begin{equation}
 g \ll  1 \ .
\label{73}
\end{equation}
On the other hand, as the comparison of  the correction (\ref{64})
with the lowest
order ground state result (\ref{37})  shows,   such an expansion involves
as a small parameter
\begin{equation}
\frac{\ell}{g^{2}L} \ll  1 \ .
\label{74}
\end{equation}
Obviously these two expansions can be made compatible only if
the coupling constant
$g$ is considered a function of the number of degrees of freedom
($L/l$). With the
 dependence
\begin{equation}
g^{2} \propto 1/\ln \left(L/\ell\right)
\label{75}
\end{equation}
expected to hold in the continuum limit,  the two requirements
(\ref{73}), (\ref{74})  are indeed compatible. It is encouraging that the
non-perturbative effects indicating confinement require a behaviour
of the coupling constant in the thermodynamic limit which is
compatible with asymptotic freedom. It is tempting to speculate that
absence of confinement in Higgs-gauge theories (e.g. SU(2) with a
scalar doublet) is in turn related to a behaviour of $g$ in
the continuum limit of such models \cite{HAHA} which, in contradistinction
to (\ref{75}), is controlled by
a cutoff rather than by the number of degrees of freedom.
\newpage

\noindent
{\bf Acknowledgements}
\vskip 0.2cm
\noindent
Stimulating discussions with  K. Johnson and A. K. Kerman
are gratefully acknowledged.
This work was supported in part by the Alexander von Humboldt
Stiftung
and by DOE under the grant number
DE-FG02-94ER40823. F. L. and M. T. are supported by the Bundesministerium
f\"{u}r Forschung und Technologie.
\vskip 2.0cm

\bibliographystyle{unsrt}

\end{document}